\documentclass[preprint,prab, nofootinbib]{revtex4-2}

\usepackage[colorlinks=true, allcolors=blue]{hyperref}
\usepackage{graphics}
\graphicspath{{images/}{}}
\usepackage{dcolumn}
\usepackage{bm}
\usepackage{siunitx}
\usepackage{xcolor}
\usepackage{braket}
\usepackage{mathtools}
\usepackage{amsmath}
\usepackage{leftidx}
\usepackage{wasysym}
\usepackage{float}
\usepackage{cleveref}
\Crefname{equation}{Eq.}{Eqs.}

\usepackage{amsfonts} 
\DeclareMathSymbol{\shortminus}{\mathbin}{AMSa}{"39}

\graphicspath{{images/}}

\begin{document}

\title{Cooling and diffusion rates in coherent electron cooling concepts}

\author{Sergei Nagaitsev}
\email{nsergei@fnal.gov}
\altaffiliation[Also at ]{the University of Chicago,
Chicago, Illinois 60637, USA}
\author{Valeri Lebedev}
\affiliation{Fermi National Accelerator Laboratory, Batavia, IL 60510, USA}%
\author{Gennady Stupakov}
\affiliation{SLAC National Accelerator Laboratory, Stanford University, Menlo
Park CA 94025, USA}
\author{Erdong Wang}
\author{William Bergan}
\affiliation{Brookhaven National Laboratory, Upton NY 11973, USA}
\date{\today}

\begin{abstract}
We present analytic cooling and diffusion rates for a simplified model of coherent electron cooling (CEC), based on a proton energy kick at each turn. This model also allows to estimate analytically the rms value of electron beam density fluctuations in the "kicker" section. Having such analytic expressions should allow for better understanding of the CEC mechanism, and for a quicker analysis and optimization of main system parameters. Our analysis is applicable to any CEC amplification mechanism, as long as the wake (kick) function is available.

\end{abstract}

\maketitle

\section{Introduction}

\begin{figure}[hb!]
    \centering
    \includegraphics[width=0.9\linewidth]{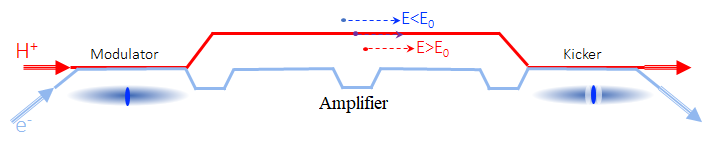}
    \caption{\label{fig:cec_schematic} A simplified schematic of CEC. }
\end{figure}

Let us consider a 1D longitudinal coherent electron cooling (CEC) scheme as proposed in Ref.~\cite{Derbenev_CeC_1992, PhysRevLett.102.114801, PhysRevLett.111.084802,Stupakov:2018anp, PhysRevAccelBeams.22.034401}. Figure \ref{fig:cec_schematic} presents a simplified schematic of CEC.  The electron bunch picks up density modulations from co-propagating protons in the "Modulator" section.  These density modulations are then amplified by some mechanism in the "Amplifier" section (blue).

The proton beam line (red) is arranged in such a way that when protons arrive at the "Kicker" section, faster (slower) protons overcome (lag behind) a reference on-energy particle.  

\begin{figure}[hb!]
    \centering
    \includegraphics[width=0.6\columnwidth]{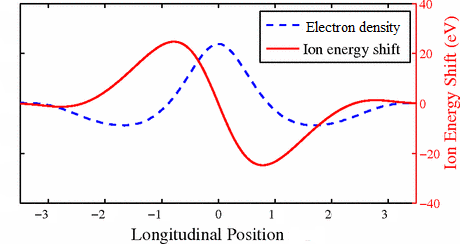}
    \caption{\label{fig:cec_kick} The electron beam density modulation due to a single proton (arb. units) and a corresponding energy kick (in eV) after the "Kicker" section as a function of the proton's longitudinal position ($\mu$m). }
\end{figure}

In our simplified model we will assume that at the end of the "Kicker" section, the proton energy experiences a kick as shown in Fig. \ref{fig:cec_kick}. For convenience, we will call the proton energy change dependence versus $z$ the \emph{wake function}---apart from a different normalization, it is the same as the conventional longitudinal wake in accelerator physics. For simplicity, we will assume that the proton's longitudinal position, $z$, in the "Kicker" section does not change and is equal to $z=R_{56} \delta$, where $\delta = \frac{\delta p}{p_0}$ is the proton's relative momentum deviation and $R_{56}$ is the proton-line linear transfer matrix element from the end of the "Modulator" section to the "Kicker" section, i.e. $z$ depends only on the proton momentum deviation.  One can now see from Fig. \ref{fig:cec_kick} that faster (slower) protons would lose (gain) energy after the "Kicker" section passage.  
The wake function, introduced above, is the main element in various modifications of coherent electron cooling.  For the microbunched electron cooling (MBEC) concept it was calculated in Refs.~\cite{PhysRevAccelBeams.22.034401}; for the plasma-cascade (PCA) cooling concept, the wake function can be found in Ref.~\cite{Wang:2019dkt}.
In what follows, we will use the wake function calculated for an MBEC cooler currently being designed for the electron-ion collider (EIC) (for details see Ref.~\cite{Bergan_2021}) and shown in Fig~\ref{fig:cec_kick}. Table \ref{tab:CEC_parameters} gives an example of system parameters, used in our calculations.
We will discuss both the cooling rate and the diffusion rate due to neighboring protons producing random kicks and, thus, creating a heating mechanism. Other diffusion mechanisms will also be considered. 

\begin{table}[htp]
\centering
\caption{ \label{tab:CEC_parameters}CEC system parameters (example)}
\begin{tabular}{|| l ||  c || r || l ||}
\hline
\textbf{Parameter} & \textbf{Symbol}  & \textbf{Value} & \textbf{Unit} \\
\hline
Proton energy & $E_0$ & 275  & GeV \\
Lorentz factor & $\gamma$ & 290  &  \\
Ring circumference & $C$ & 3834  & m \\
Revolution frequency & $f_0$ & 78.3  & kHz \\
Number of protons per bunch & $N_p$ & 6.9  & $10^{10}$  \\
Proton rms momentum spread & $\delta_p$ & 6.8  & $10^{-4}$ \\
Proton rms bunch length & $\sigma_{pz}$ & 6.0  & cm \\
\hline
Number of electrons per bunch & $N_e$ & 6.3  & $10^{9}$  \\
Electron rms bunch length & $\sigma_{ez}$ & 4.0  & mm \\
Electron rms beam size (vertical) & $\sigma_{ey}$ & 0.6  & mm \\
Electron rms beam size (horizontal) & $\sigma_{ex}$ & 0.6  & mm \\
Kicker section length & $L_k$ & 40  & m \\\hline
\end{tabular}
\end{table}

\section{Energy kick}

\begin{figure}[h!]
    \centering
    \includegraphics[width=0.9\columnwidth]{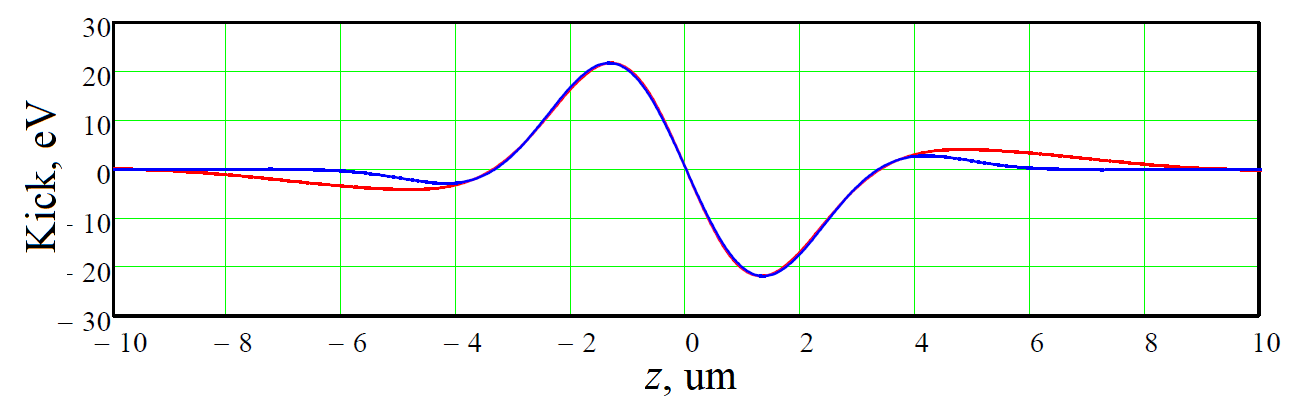}
    \caption{\label{fig:cec_kick_model} The energy kick (eV) after the "Kicker" section as a function of the proton's longitudinal position $z = R_{56} x$ ($\mu$m) with respect to the reference on-energy proton. The red curve is a calculated wake, based on Ref. \cite[Eq. C7]{transv_cooling_MBEC}. The blue curve is the proposed approximation, Eq. (\ref{eq:kick}). }
\end{figure}

To allow for analytical treatment of the problem, we will use the following model expression for the proton energy kick in the "Kicker" section,
\begin{equation}\label{eq:kick}
    w(z) = - V_0 \sin{ \left ( 2\pi \frac{z}{z_0} \right )} \exp{\left ( - \frac{z^2}{{\sigma_0}^2} \right )}, 
\end{equation}
where we introduced three adjustable parameters: $V_0$, the amplitude of the kick, $z_0$, the characteristic wavelength, and $\sigma_0$, the characteristic width. The negative sign reflects the fact that the leading particle ($z>0$) loses its energy after the kick. For example, the energy kick, calculated using the system parameters in Table \ref{tab:CEC_parameters} and shown in Fig.~\ref{fig:cec_kick}, is presented in Figure \ref{fig:cec_kick_model} (red curve) together with our model, Eq. (\ref{eq:kick}) (blue curve).  
One can see from Figure \ref{fig:cec_kick_model} that the proposed approximation slightly underestimates the far tales of the wake.  This does not affect the cooling rate but slightly underestimates the diffusion rate.

For the calculated energy kick, the following model parameters provide the best fit: $V_0 = 28$ eV, $z_0 = 6.7$ $\mu$m, and $\sigma_0$ = 3.0 $\mu$m. One can notice that at $|z| > \frac{z_0}{2}$ the energy kick changes its sign and cooling becomes anti-cooling.  This determines the so-called cooling range, the number of "sigmas" $n$ such that $n R_{56} \delta_p = {z_0}/{2}$.

\section{Fokker-Planck equation}

To describe the evolution of the proton momentum distribution function, we will use the Fokker-Planck equation in the following form:
\begin{equation}\label{eq:fokker-planck}
    \frac{\partial \psi}{\partial t} + \Dot{s} \frac{\partial \psi}{\partial s} + \Dot{\delta} \frac{\partial \psi}{\partial \delta} = - \frac{\partial}{\partial \delta} \left ( F(\delta, s)\psi  \right ) + \frac{1}{2} \frac{\partial}{\partial \delta} \left ( D(\delta, s) \frac{\partial \psi }{\partial \delta}  \right ), 
\end{equation}
where $\psi(\delta, s, t)$ is the proton distribution function, $\delta$ is the relative momentum deviation, $s$ is the longitudinal coordinate in the lab frame (with respect to the bunch center), $F(\delta, s) = f_0 w(R_{56} \delta, s)/E_0$ is the cooling force 
and $D(\delta, s)$ is the diffusion coefficient. The diffusion can include various contributions, such as heating due to near-by protons, electron beam noise, intra-beam scattering, etc. Eq.~(\ref{eq:fokker-planck}) corresponds to a bunched-beam case. We will transform Eq.~(\ref{eq:fokker-planck}) to unperturbed longitudinal action-angle variables $(J, \phi)$ in order to analyse the cooling and diffusion processes in terms of the longitudinal bunch emittance~\cite{Ruggiero:1993uj, bax_stu_2020},
\begin{equation}\label{eq:action-angle}
    \delta = \sqrt{\frac{2 J}{\beta}} \sin{\phi}, \quad s = \sqrt{2 J \beta} \cos{\phi},
\end{equation}
where $\beta$ is the so-called longitudinal beta function, $\beta=\sigma_{pz}/\delta_p \approx$ 88 $\si{m}$ for the parameters in Table \ref{tab:CEC_parameters}.
If the characteristic cooling and diffusion times are longer than the synchrotron oscillation period, it is reasonable to assume that the bunch distribution is continuously matched to the shape of the trajectories in the ($J,\, \phi$)-phase plane  and that the distribution function depends explicitly only on $J$ and not on $\phi$, that is $\psi = \psi(J,\, t)$.  This simplifies considerably the left-hand side of Eq. (\ref{eq:fokker-planck}), 
\begin{equation}\label{eq:fokker-planck-J}
    \frac{\partial \psi}{\partial t}  = - \sqrt{2 \beta} \frac{\partial}{\partial J} \left ( \sqrt{J} \Tilde{F}(J) \psi  \right ) + \beta \frac{\partial}{\partial J} \left ( J \Tilde{D}(J) \frac{\partial \psi }{\partial J}  \right ), 
\end{equation}
where the cooling force $\Tilde{F}$ is given by
\begin{equation}\label{eq:cooling-force-J}
    \Tilde{F}(J) = \frac{1}{2 \pi} \int_0^{2 \pi} F(\delta, s) \, \sin{\phi} \, d \phi
\end{equation}
and the diffusion term $\Tilde{D}$ is given by
\begin{equation}\label{eq:diffusion-J}
    \Tilde{D}(J) = \frac{1}{2 \pi} \int_0^{2 \pi} D(\delta, s) \, \sin^2{(\phi)} \, d \phi
\end{equation}
with $\delta$ and $s$ given by Eq. (\ref{eq:action-angle}). For a detailed derivation of Eqs. (\ref{eq:cooling-force-J}) and (\ref{eq:diffusion-J}), see Ref. \cite{bax_stu_2020}. 

In its simplest form, the cooling force can be presented as $F(\delta, s) = - \lambda \delta$, while the diffusion as a constant $D(\delta, s) = D_0$. The Fokker-Planck equation becomes 
\begin{equation}\label{eq:fokker-planck2}
    \frac{\partial \psi  }{\partial t} 
    - 
    \lambda \frac{\partial}{\partial J} \left ( J\,\psi  \right ) 
    = 
    \frac{\beta \, D_0}{2}  
    \frac{\partial }{\partial J}
    \left ( J
    \frac{\partial \psi  }{\partial J} 
    \right ). 
\end{equation}

We will now multiply both sides of Eq. (\ref{eq:fokker-planck2}) by $J$ and integrate  in order to obtain the evolution of the rms longitudinal emittance, $\epsilon_L = \int_0^\infty \psi \, J dJ$ (here we assume that $\psi$ is normalized by unity, $\int_0^\infty \psi \, dJ=1$),
\begin{equation}\label{eq:rms-evolution}
   \frac{d \epsilon_L}{d t} + \lambda \, \epsilon_L = \frac{\beta \, D_0}{2}. 
\end{equation}
In a steady state (that is for $d/dt = 0$), the equilibrium rms emittance is 
\begin{equation}\label{eq:rms-equilib}
   \epsilon_L = \frac{D_0 \, \beta}{2\lambda}. 
\end{equation}
For a more realistic cooling force, we notice (see Table \ref{tab:CEC_parameters}) that the electron bunch is much shorter than the proton bunch, $\sigma_{ez} \ll \sigma_{pz}$. We will therefore use the following cooling force approximation:
\begin{equation}\label{eq:cooling-force-ps}
    F(\delta, \,s) = - \frac{f_0 V_0}{E_0} \sin{ \left ( 2\pi \frac{R_{56} \delta}{z_0} \right )} \exp{\left ( - \frac{R_{56}^2 \delta^2}{{\sigma_0}^2} \right )} \exp{\left (- \frac{s^2}{\sigma_{ez}^2} \right )} 
\end{equation}
with $\delta$ and $s$ given by Eq. (\ref{eq:action-angle}).
This equation assumes that the electron bunch is placed at the center of the proton bunch and the interaction happens only for protons with $s \approx 0$, because $\sigma_{ez} \ll \sigma_{pz}$. Therefore, we can use the following approximation: $\exp{\left (- \frac{s^2}{\sigma_{ez}^2} \right )} \approx \sqrt{\pi} \sigma_{ez} \delta(s)$, where $\delta(s)$ is the Dirac delta function.
Using Eq. (\ref{eq:cooling-force-J}), one can obtain the cooling force as a function of action $J$: 
\begin{equation}\label{eq:cooling-force-J1}
    \Tilde{F}(J) = - \frac{f_0 V_0}{E_0} \frac{\sigma_{ez}}{\sqrt{2 \pi \beta J}} \sin{ \left ( 2\pi \frac{R_{56}}{z_0} \sqrt{\frac{2 J}{\beta}} \right )} \exp{\left ( - \frac{R_{56}^2 }{{\sigma_0}^2} \frac{2 J}{\beta} \right )}. 
\end{equation}

\section{Cooling rate}

\begin{figure}[h!]
    \centering
    \includegraphics[width=0.9\columnwidth]{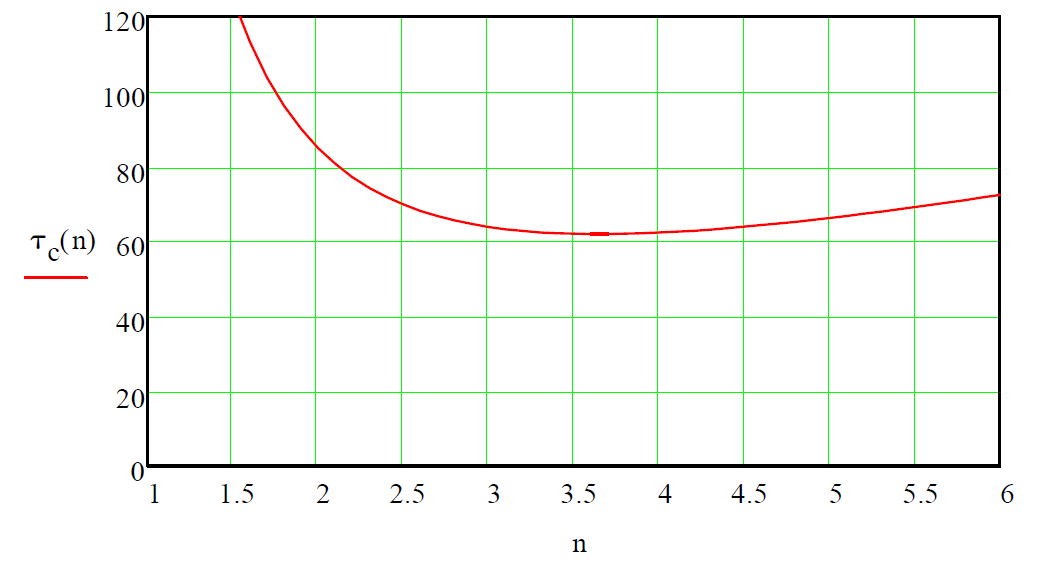}
    \caption{\label{fig:cec_cooling_time1} Cooling time (in minutes) as a function of the cooling range $n$ defined as $n R_{56} \delta_p = \frac{z_0}{2}$.  }
\end{figure}

To obtain the cooling rate, $\tau_c$, from the fokker-Plank equation, Eq. (\ref{eq:fokker-planck-J}), we will evaluate the following integral:
\begin{equation}\label{eq:cooling-rate2}
    \frac{1}{\tau_c} = \frac{\sqrt{2 \beta}}{\epsilon_L}
     \int_{0}^{+\infty} J \frac{\partial}{\partial J} \left ( \sqrt{J} \Tilde{F}(J) \psi \right ) dJ,
\end{equation}
where $\Tilde{F}(J)$ is given by Eq. (\ref{eq:cooling-force-J1}) and the distribution function
\begin{equation}\label{eq:dist-function}
    \psi = \frac{1}{\epsilon_L} \exp{\left ( - \frac{J}{\epsilon_L} \right )}. 
\end{equation}
The resulting cooling rate is
\begin{equation}\label{eq:cooling-rate3}
    \frac{1}{\tau_c} = \frac{ \pi f_0 V_0 }{\delta_p n E_0} \frac{\sigma_{ez}}{\sqrt{2} \sigma_{pz}} \left ( 1 + \frac{{z_0}^2}{2 n^2 {\sigma_0}^2} \right )^{-3/2} \exp{\left ( - \frac{\pi^2}{2n^2 + {{z_0}^2}/{{\sigma_0}^2} } \right )} 
     ,
\end{equation}
where $n$ is the cooling range, such that $n R_{56} \delta_p = {z_0}/{2}$.  Figure \ref{fig:cec_cooling_time1} shows the cooling time, $\tau_c$, as a function of the cooling range, $n$, for the CEC system parameters in Table \ref{tab:CEC_parameters} and $V_0 = 28$ eV, $z_0 = 6.7$ $\mu$m, and $\sigma_0$ = 3.0 $\mu$m.
One can see that there is a shallow minimum of about 60 minutes for $n$ in the range 3.5 to 4.5. For example, choosing $n$ = 3.7 results in $R_{56} = Z_0/(2n \delta_p) \approx$ 1.3 mm. This value should be compared to the kinematic portion of the $R_{56}$ element. If the proton path length between the "Modulator" and the "Kicker" section is $L \approx$ 100 m, the kinematic portion of the $R_{56}$ element is $L/\gamma^2 \approx$ 1.2 mm.  Thus, the proton beam line has to provide an additional 0.1 mm increase to the $R_{56}$ matrix element.  It also means that the Kicker section cannot be too long as its length\footnote{There are additional constraints on the Kicker section length, due to plasma oscillations in the electron beam, for example.} increases the effective value of the $R_{56}$ element. For $n \to \infty$, the cooling time increases linearly with $n$ and the Eq. (\ref{eq:cooling-rate3}) becomes: ${\tau_c}^{-1} \approx \lambda$, as expected.

\section{Diffusion rate}

The diffusion coefficient, $D(\delta, s)$, is usually a function of the momentum deviation, $\delta$. However, it was shown in Ref.~\cite{lebedev2020stochastic} that in the case of stochastic cooling with a strong Schottky band overlap, the diffusion coefficient due to random kicks from neighbouring protons is a constant, i.e. independent of $\delta$. The CEC method, having the typical frequencies of $c/z_0 \approx 45$ THz, is in the regime of a strong Schottky band overlap.  In this regime, the diffusion coefficient at the center of the electron bunch can be written as 
    \begin{align}
        D_0
        =
        \frac{\langle (w(z)/E_0)^2\rangle}{T}
        ,
    \end{align}
where the angular brackets $\langle ... \rangle $ indicate averaging of random energy kicks from neighboring ions, and $T=1/f_0$ is the revolution period in the ring. Taking into account that the number of ions per unit length at the center of a Gaussian bunch is $N_p/(\sqrt{2\pi}\sigma_{pz})$, we obtain
\begin{equation}\label{eq:diff-rate2}
    D_0 = f_0 \frac{N_p}{\sqrt{2 \pi} \sigma_{pz}} \frac{{V_0}^2}{{E_0}^2} \int_{-\infty}^{+\infty} \left ( \sin{ \left ( 2\pi \frac{z}{z_0} \right )} \exp{\left ( - \frac{z^2}{{\sigma_0}^2} \right )} \right )^2 dz.
\end{equation}
We can finally write
\begin{equation}\label{eq:diff-rate3}
    D_0 = \frac{N_p f_0 {V_0}^2}{4 {E_0}^2}
    \frac{\sigma_0}{\sigma_{pz}}
    \left ( 1 - \exp{\left ( - 2 \pi^2 \frac{{\sigma_0}^2}{{z_0}^2} \right )} \right ) \approx \frac{N_p f_0 {V_0}^2}{4 {E_0}^2}
    \frac{\sigma_0}{\sigma_{pz}}.
\end{equation}
As expected, the diffusion rate is independent of the cooling range $n$ and is proportional to the width of the kick, $\sigma_0$, which can be viewed as the inverse band-width of the system. Recalling that the electron bunch is much shorter than the proton bunch, Eq. (\ref{eq:cooling-force-ps}), we can write the diffusion coefficient for any longitudinal position, $s$, within the proton bunch as
\begin{equation}
    D(s) = D_0 \exp{\left ( -\frac{2 s^2}{\sigma_{ez}^2} \right )} \approx 
    D_0 \sqrt{\frac{\pi}{2}} \sigma_{ez} \delta(s).
\end{equation}
After averaging over angle $\phi$ by using Eq. (\ref{eq:diffusion-J}) we obtain
\begin{equation} \label{eq:diff-rate-J1}
    \Tilde{D} (J) = \frac{D_0}{2 \sqrt{\pi}} \frac{\sigma_{ez}}{\sqrt{\beta J}}.
\end{equation}
From Eq. (\ref{eq:fokker-planck-J}) the evolution of the longitudinal rms emittance, $\epsilon_L$ is determined by
\begin{equation}\label{eq:rms-evolution1}
   \frac{1}{\epsilon_L} \frac{d {\epsilon_L}}{d t} = - \frac{1}{\tau_c} + \frac{D_0 \sigma_{ez} \sqrt{\beta}}{4 \sqrt{\epsilon_L^3}}, 
\end{equation}
    \noindent with $\tau_c$ from Eq. (\ref{eq:cooling-rate3}) and we assumed that $\sigma_{pz} = \sqrt{\beta \epsilon_L}$. The Eq. (\ref{eq:rms-evolution1}) is valid for $\sigma_{ez} \ll \sigma_{pz}$.
For the CEC system parameters in Table \ref{tab:CEC_parameters} and for $V_0 = 28$ eV, $z_0 = 6.7$ $\mu$m, and $\sigma_0$ = 3.0 $\mu$m, the diffusion time is  $\approx$ 660 minutes, which is much greater than the cooling time for the same parameters and $n$ = 3.7, $\tau_c \approx$ 62 minutes.  This indicates that, in theory, the overall sum of cooling and diffusion rates in Eq. (\ref{eq:rms-evolution1}) can still be increased by increasing the kick amplitude, $V_0$. For the so-called "optimal gain" \cite{MOHL198073} condition we have:
\begin{equation}\label{eq:opt-gain}
   \frac{1}{\tau_c} = \frac{D_0 \sigma_{ez} \sqrt{\beta}}{2 \sqrt{\epsilon_L^3}}. 
\end{equation}
From this we can obtain the optimal kick amplitude, $V_{opt}$, for maximum cooling:
\begin{equation}\label{eq:opt-kick}
    V_{opt} = \frac{4 \sqrt{2} \pi E_0 \delta_p }{n N_p} \frac{\sigma_{pz}}{ \sigma_{0}} 
    \left ( 1 + \frac{{z_0}^2}{2 n^2 {\sigma_0}^2} \right )^{-3/2} \exp{\left ( - \frac{\pi^2}{2n^2 + {{z_0}^2}/{{\sigma_0}^2} } \right )}.
\end{equation}
This yields $V_{opt} \approx$ 150 eV. With this optimal kick amplitude, the achievable cooling time becomes $2 \tau_c \approx$ 24 minutes (the factor of 2 is due to Eq. (\ref{eq:opt-gain})). We note that only the diffusion due to neighboring protons is taken into account in Eq. (\ref{eq:opt-kick}).  Other diffusion mechanisms can be added to analyze the effective cooling rate in Eq. (\ref{eq:rms-evolution1}).   

\section{Electron beam density fluctuations due to protons}

First, we can estimate the rms value of random energy kicks per turn due to diffusion Eq. (\ref{eq:diff-rate3}) for a proton at the center of the electron bunch.  For the CEC parameters Table \ref{tab:CEC_parameters}, the diffusion coefficient $D_0 \approx 2.3 \times 10^{-11}$ sec$^{-1}$. The rms energy kick per one turn due to diffusion can be written as
\begin{equation}\label{eq:rms-kick1}
   \delta E_{rms} = E_0 \sqrt{D_0  T} \approx 26~\si{keV}. 
\end{equation}
This random kick is due to other protons in the vicinity of a reference proton.  This should be compared to a cooling wake (kick) value of about 20 V (max) per turn as can be seen in Fig. (\ref{fig:cec_kick_model}).

Let us estimate the rms electron beam density fluctuations, resulting from the superposition of incoherently-added wakes from neighbouring protons. First, we will use a simplified single-wavelength density modulation model for a single proton with $k=2 \pi / z_0$, a wave vector of this modulation.  In the electron rest-frame, we can use the Poisson's electrostatics equation (in Gaussian units):
\begin{equation}\label{eq:Poisson1}
   \frac{d E_z}{d z'} = 4 \pi e n_e(z'), 
\end{equation}
where $e$ is the electron charge and $z'$ is the transformed rest-frame $z$ coordinate (note that $E_z = {E'}_z$). Using Eq.~\eqref{eq:Poisson1} for the calculation of the electron density perturbation we actually replace electrons by uniformly charged thin slices and assume that the distance between the slices is much smaller than their transverse size.  A more accurate model of Gaussian slices with elliptic cross-section and an arbitrary transverse size was used in Ref.~\cite{transv_cooling_MBEC}.

From Eq.~\eqref{eq:Poisson1} we find the following electron density modulation amplitude (in the beam rest frame) due to a single proton:
\begin{equation}\label{eq:Poisson2}
   {n'}_e = \frac{k E_z}{4 \pi \gamma e} . 
\end{equation}
The rest-frame electric field $E'_z$ can be estimated as $E'_z = E_z \approx V_0/(e L_k)$. Thus, we obtain for a single proton, the amplitude of the density modulation,
\begin{equation}\label{eq:Poisson3}
   {n'}_e \approx \frac{k V_0}{4 \pi \gamma e^2 L_k} . 
\end{equation}
The average over $z'$ of this density modulation is zero, $\langle {n'}_e \rangle = 0$, but the rms value is non-zero:
\begin{equation}\label{eq:density1}
   \sqrt {\langle {n'}_e^2 \rangle} \approx \frac{k V_0}{4 \pi \gamma e^2 L_k} \sqrt{\Delta N_p}, 
\end{equation}
where $\Delta N_p$ is the number of protons in a sample, near a reference proton:
\begin{equation}\label{eq:density2}
   \Delta N_p = \frac{N_p \sigma_0}{\sqrt{2 \pi} \sigma_{pz}}. 
\end{equation}
We finally arrive at an estimate for the rms electron beam density modulation:
\begin{equation}\label{eq:density3}
   \sqrt {\langle {n'}_e^2 \rangle} \approx \frac{k V_0}{4 \pi \gamma e^2 L_k} \sqrt{\frac{N_p \sigma_0}{\sqrt{2 \pi} \sigma_{pz}}}.
\end{equation}
The maximum electron bunch density in the rest-frame is
\begin{equation}\label{eq:density4}
   n'_{e0} = \frac{N_e}{\sqrt{(2 \pi)^3} \gamma \sigma_{ez} \sigma_{ex} \sigma_{ey}}.
\end{equation}
For the parameters of Table \ref{tab:CEC_parameters} we obtain,
\begin{equation}\label{eq:density_ratio}
   \frac{\sqrt {\langle {n'}_e^2 \rangle}}{n'_{e0}} \approx \frac{k V_0 \sigma_{ez} \sigma_{ex} \sigma_{ey}}{2 e^2 L_k} \frac{\sqrt{N_p}}{N_e} \sqrt{\frac{\sqrt{2 \pi} \sigma_0}{ \sigma_{pz}}} \approx 0.15.
\end{equation}
Thus, a simplified estimate gives a 15\% relative rms density fluctuations value in the electron bunch, needed to support cooling in the presence of other protons\footnote{Eq.~\eqref{eq:Poisson1} treats the particles as charged sheets and thus overestimates their interaction at large distance. A more accurate model of elliptic slices from Ref.~\cite{transv_cooling_MBEC} takes into account that the interaction is localized at the distance $\sim\sigma_{ex}/\gamma,\sigma_{ey}/\gamma$; it gives a larger value for the estimate of ${\sqrt {\langle {n'}_e^2 \rangle}}/{n'_{e0}}$ that depends on the cross section of the electron and proton beams.}. Obviously, it's not a small number and may require a separate investigation.  The maximum possible modulation level can not exceed 100\% and, thus, the maximum rms value should probably be limited to 30\% to allow for variations in excess of 2-3 sigmas.  One might also notice that achieving the optimal gain of $V_0 \approx 150$ eV may not be possible for the parameters of Table \ref{tab:CEC_parameters}.
A more detailed analysis, obtained by performing averaging of $n_e^2(z')$ with 
\begin{equation}\label{eq:density5}
   n_e(z') = \frac{1}{4 \pi e^2 L_k} \frac{d w(z')}{d z'},
\end{equation}
yields 
\begin{align}
\label{eq:density6}
\begin{split}
   \sqrt {\langle {n'}_e^2 \rangle} & = \sqrt{\frac{N_p}{\sqrt{2 \pi} \gamma \sigma_{pz}}\int^{+\infty}_{-\infty} n_e^2(z') dz'} 
   \\
   & = \frac{V_0}{4 \pi \gamma e^2 L_k} \sqrt{\frac{N_p}{4 \sigma_0 \sigma_{pz}}} \sqrt{k^2 \sigma_0^2 + 1 - \exp{ \left (- \frac{k^2 \sigma_0^2}{2}\right)}}.
\end{split}
\end{align}
Using Eq. (\ref{eq:density6}) yields the relative rms density fluctuation of 13\%.

\section{Electron beam noise}
In this section we estimate the electron beam shot-noise contribution to the diffusion coefficient. Since the electron longitudinal charge density is similar to that of protons (see Table \ref{tab:CEC_parameters}), we can estimate the electron shot-noise contribution to the diffusion to be similar to the proton beam contribution, Eq. (\ref{eq:diff-rate3}).  This doubles the effective diffusion coefficient and  gives the effective diffusion time of $(\Tilde{D} \beta/{\epsilon_L})^{-1} \approx$ 330 minutes, still much greater than the cooling time, $\approx 60$ minutes.  One can see that exceeding the shot-noise value in the electron beam by a factor of 2-3 is possible for the chosen parameters, before the electron beam noise becomes a dominant diffusion factor.  One has to remember, however, that doubling the diffusion coefficient increases the rms electron beam density fluctuations by a factor of $\sqrt{2}$, to over 20\%.  Thus, this may also become a limiting factor.
For a more detailed calculation of electron beam contributions to the diffusion rate, one needs to use the spectral power density of the electron beam density fluctuations and then take into account the amplification section to convert these density fluctuations to electric fields.  This requires a separate investigation, outside of the scope of this note.

\section{Conclusions}
In this note, we considered a simplified cooling wake (kick) model, given by Eq. (\ref{eq:kick}). This model allows to derive analytic expressions for cooling and diffusion rates, as well as for the electron beam rms density fluctuations.  Having such analytic expressions should allow for better understanding of the CEC mechanism, and for a quicker analysis and optimization of main system parameters.

We would like to emphasize that even though we have used the wake calculated for the MBEC amplification scheme, our analysis can be easily applied to other coherent cooling techniques, for example, to the PCA concept, as long as the wake (kick) function, similar to the one shown in Fig.~\ref{fig:cec_kick_model}, is available.

\section{Acknowledgements}
This manuscript has been authored by Fermi Research Alliance, LLC under Contract No. DE-AC02-07CH11359 with the U.S. Department of Energy, Office of Science, Office of High Energy Physics. This work was also supported by the Department of Energy, Contract No. DE-AC02-76SF00515 and by Brookhaven Science Associates, LLC under Contract No. DE-SC0012704.

\bibliography{bibliography} 

\end{document}